\documentclass[pra,twocolumn,amsmath]{revtex4}

\usepackage{graphicx}
\usepackage{dcolumn}
\usepackage{bm}

\begin{document}

\title{Testing the Fair Sampling Assumption for EPR-Bell Experiments with Polarizing Beamsplitters}

\author{Guillaume Adenier}
 \altaffiliation{Pre-doc fellowship of EU-Network ``QP and Applications"}
 \email{guillaume.adenier@msi.vxu.se}
\author{Andrei Yu. Khrennikov}%
 \email{andrei.khrennikov@msi.vxu.se}
\affiliation{%
Department of Mathematics and System Engineering, V\"axj\"o
University, 351 95 V\"axj\"o, Sweden.
}%

\date{\today}

\begin{abstract}
In spite of many attempts, no local realistic model seems to be
able to reproduce EPR-Bell type correlations, unless non ideal
detection is allowed. The low efficiency of detectors in all
experiments with photons makes the use of the fair sampling
assumption unavoidable. However, since this very assumption is
false in all existing local realistic models based on inefficient
detection, we thus question its validity. We show that it is no
more reasonable to assume fair sampling than it is impossible to
test, and we actually propose an experimental test which would
provides clear cut results in case of unfair sampling.
\end{abstract}

\maketitle

\section{\label{sec:level1}Introduction}
The role of the efficiency of detectors in EPR-Bohm experiments
\cite{epr,bohm} with photons has been the subject of continuous
investigation during the last thirty years, since allowing for non
ideal detection is the only acknowledged way to get an apparent
violation of CHSH-Bell Inequality \cite{bell64,CHSH} with a local
realistic model \footnote{Some references on recent attempts to
explain EPR-Bohm type correlations with local realistic models
based on other loopholes than efficiency can be found in
\cite{Qtrf}.}. The main difference between existing experiments
\footnote{A recent experiment with massive particles \cite{Rowe}
did reach perfect detection efficiency, but at the cost of
reopening the locality loophole.} and local realistic models thus
based on inefficiency is the status of the {\itshape fair
sampling} assumption. On one hand, in order to assess that a
violation of Bell inequality has occurred, all EPR-Bell
experiments with photons are interpreted assuming fair sampling,
as it is taken to be both a reasonable and experimentally
untestable assumption. On the other hand, this very assumption is
precisely false in all existing local realistic models based on
inefficiency. Yet, these models are supposed to be only {\itshape
ad hoc}, designed to show that a violation of Bell inequality is
possible in principle, but usually without any claim for relevance
with physical reality.

The purpose of this article is to question this fair sampling
assumption. We will show that a very straightforward model using
contextual probabilities \cite{Khrennikov1} can reproduce EPR
correlations and that it is possible to overcome the usual
understanding of the fair sampling issue with a generic test
capable of disproving the fair sampling assumption.

\section{\label{sec:level2}A simple local hidden-variable model based on the detection loophole}
\begin{figure}
\center
\includegraphics[width=7.0cm]{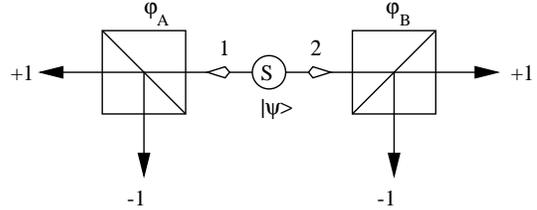}
\caption{\label{fig:epsart8} Two-channel EPR-Bell experiment with
photons. The source is the rotationally invariant singlet state
$|\psi\rangle$ and the measurement is made by two polarizing
beamsplitters with orientation settings $\varphi_A$ and
$\varphi_B$, monitored by a fourfold coincidence setup (not
represented here).}
\end{figure}
The well-known scheme for an EPR-Bell experiment is represented on
Fig.~\ref{fig:epsart8}. A source sends pairs of photons, labelled
1 and 2, in an entangled singlet state $|\psi\rangle$, and
polarization measurements are carried out on these photons using
polarizing beam splitters (PBS) A and B oriented respectively
along $\varphi_A$ and $\varphi_B$. The possible measurement
results for each particle are labelled as +1 if the particle is
detected in the ordinary channel, and as -1 if it is detected in
the extraordinary channel.

The rates of the four possible coincident events are labelled
\footnote{These rates depends on the experimental settings
$\varphi_A$ and $\varphi_B$ and on the polarization distribution
$\rho$ of the source. This dependence is left here implicit for
simplicity.} $R_{++}$, $R_{+-}$, $R_{-+}$, and $R_{--}$. For
instance $R_{+-}$ is the rate of coincident detection in the +1
channel of polarizing beamsplitter A and in the -1 channel of
polarizing beamsplitter B. Let $R_{d}$ be the total rate of
detected pairs, defined as:
\begin{equation}\label{totalrate}
    R_{d}(\rho,\varphi_A,\varphi_B)=R_{++}+R_{+-}+R_{-+}+R_{--}.
\end{equation}
It is then possible to define the correlation function as
\begin{equation}\label{correl}
    E(\rho,\varphi_A,\varphi_B)=\frac{R_{++}-R_{+-}-R_{-+}+R_{--}}{R_{d}(\rho,\varphi_A,\varphi_B)}
\end{equation}
As is well known, experiments with pairs of photons
\cite{Aspect82,Weihs} show a $\cos[2(\varphi_B-\varphi_A)]$
correlation, in agreement with the predictions of quantum
mechanics.

There are not many ways to get such an EPR-like correlation using
a local realistic model based on the detection loophole. This
point was already stressed by Clauser and Shimony \cite{ClauserS},
although they pointed this as a case against local
hidden-variables models, whereas for us it grants the generality
of our point. The model we will use here is indeed a simple
variant of the one that has been used many times to show
explicitly that a local realistic model based on detection
loophole can be in agreement with the experimental results
\cite{Pearle,Santos,Gisin,Larsson2,thompson}, and that these
models need not even be ``highly artificial" \cite{ClauserS}, but
can in fact be sound consequences of known principles of physics
\cite{Hofer,Kracklauer,Sanctuary}. In our model, each particle is
provided with an internal parameter \(\lambda \)---a
polarization---that can take any value in the interval \([0,2\pi
]\), and the particles issued from one pair have the same
polarization. The fate of a particle incoming into an analyzer
orientated along the direction \(\varphi \) is determined by the
difference \(\alpha =|\lambda -\varphi |\) :

\begin{itemize}
    \item If \(\alpha \) is close to 0 modulo \(\pi \)
    the particle goes into the channel labelled \(+1\).
    \item If \(\alpha \) is close to \(\pi /2\) modulo \(\pi \),
    the particle goes into the channel labelled \(-1\).
\end{itemize}

As such, the model is incapable of providing anything better than
the saw tooth, however we define ``close" and whatever the
polarization distribution of the pairs of particles (a consequence
of Bell's Theorem). However, if there exists a third channel,
labelled 0, corresponding to a non detection, measurements which
tend to reduce the curve to a saw tooth can be discarded, thus
yielding to a \(S\) greater than 2, up to 4. The particles that
must remain undetected for this purpose are the ones for which
\(\alpha \simeq \pi /4\) modulo $\pi /2$, while all other sampling
of undetected particles would reduce the correlation curve to a
saw tooth, resulting in no apparent violation of Bell Inequalities
\footnote{The rejection process of the particles is a completely
local process. It mimics nonlocality only because of the necessary
coincidence circuitry: each time one particle remains undetected
at one location, the \emph{whole} pair is \emph{logically}
rejected as providing no correlation measurement.}. We may refer
conveniently thereafter to the particles for which \(\alpha \simeq
\pi /4\) modulo $\pi /2$ as {\itshape shaky} particles. By this
terminology, we emphasized that a small perturbation of the
internal polarization \(\lambda \) of a shaky particle would
induce a change in the channel the particle would choose if it was
to be detected \footnote{This shakiness is not a property of a
particle in itself since it depends on the relative angle \(\alpha
\): it is but a quick way to label a class of particles in the
context of a given analyzer.}.

This sampling process presents the crucial feature of being
\emph{unfair}. A sampling process is thus said to be unfair if the
probability $P^r_{unfair}$ for a particle to be rejected depends
on its hidden parameters (i.e., the internal polarization
$\lambda$ in our model) and on the measurement settings (i.e., the
orientation $\varphi$ of the PBS). One can write this dependence
explicitly by stating that the probability of non detection is of
the form:
\begin{equation}\label{unfairsampling}
    P^r_{unfair}=P^r_{unfair}(\lambda,\varphi).
\end{equation}

A direct consequence of this local unfair sampling is that for
each pair of measurement settings $\varphi_1$ and $\varphi_2$ the
ensemble of detected pairs $S_{\varphi_1\varphi_2}$ belongs to a
specific probability spaces $\Omega_{\varphi_1\varphi_2}$ which
does depend on the measurement settings $\varphi_1$ and
$\varphi_2$. For each $\varphi_1$ and $\varphi_2$, some specific
regions of the Kolmogorov space are simply never recorded: the
probability space has become \emph{contextual}. In other words, a
straightforward consequence of unfair sampling is the
\emph{contextuality} of the associated probabilities
\cite{Khrennikov1}. Bell's theorem, which is derived by using one
fixed space \cite{KHR3}, is therefore no longer valid here, and
multi-context framework generalizations of Bell and CHSH
inequalities are required \cite{KHR4a,KHR4b,KHR3}.

\begin{figure}
\center
\includegraphics[width=7.0cm]{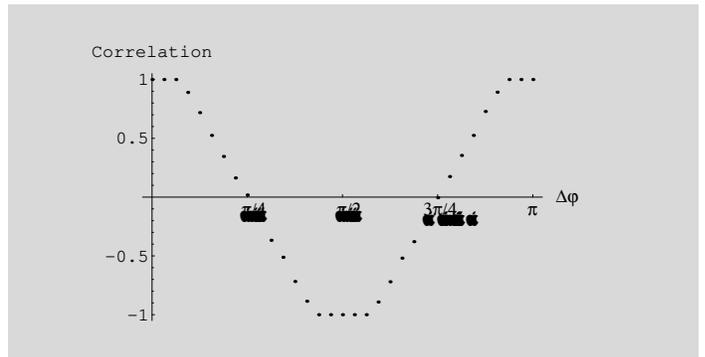}
\caption{\label{fig:epsart1} Unrealistic correlation. The internal
polarization is exactly the same for the particles from one pair.}
\end{figure}

\begin{figure}
\center
\includegraphics[width=7.0cm]{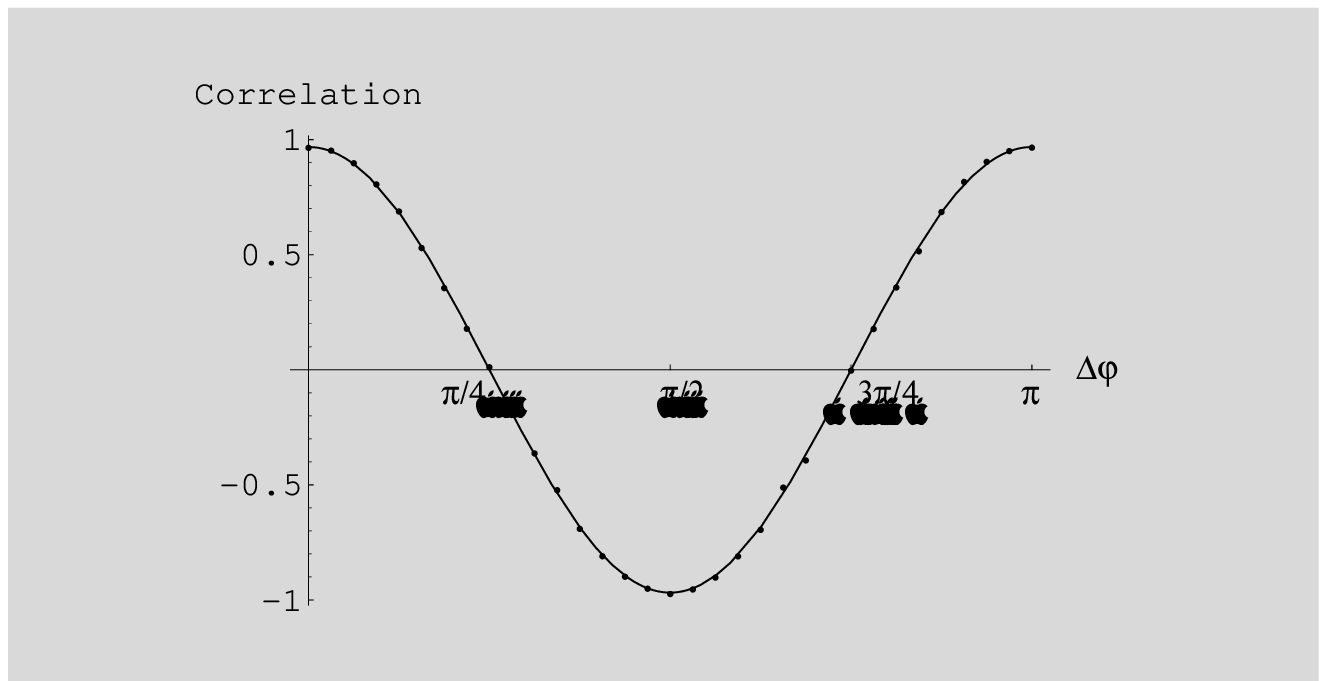}
\caption{\label{fig:epsart3} Realistic correlation. The internal
polarization differs slightly for the particles from one pair. The
result is an apparent raw violation of CHSH-Bell Inequality by
\(S=2.76\), for less than \(30\%\) pairs rejected and a visibility
of 0.97. The solid curve is not a fit to the data but the
prediction of Quantum Mechanics.}
\end{figure}

We performed a first numerical simulation \footnote{The initial
number of pairs is 10000 for each point. All simulations described
in this paper are available upon request to G. Adenier.} in the
case where both particles from one pair do share exactly the same
polarization \(\lambda \). The correlation obtained in this case
shows unrealistic sharp edges (see Fig.~\ref{fig:epsart1}).
This unrealistic correlation function can be smoothed very near a
cosinus simply by slightly breaking the statistical alignment of
the polarization of particles in accordance with a gaussian
distribution centered on perfect alignment\footnote{As a matter of
detail, we set the standard deviation of the gaussian distribution
to $\pi /16.80$, and the size of each of the four shaky regions
was $\pi /13.39$.}. The correlation then shows a good agreement
with the predictions of Quantum Mechanics (see
Fig.~\ref{fig:epsart3}).
Note that this was obtained here with unsharp polarization
correlations of particles \footnote{Contrary to the usual
understanding of entanglement as being a stronger correlation than
would allow classical physics, the use of the detection loophole
allows for loosely correlated particles to reproduce the EPR-Bohm
statistics. In our model, they actually do the job better than
strongly correlated particles.} and sharp boundaries for the
channels in the analyzer. It should however be possible to proceed
the other way round and get the same result.

\section{\label{sec:level3}Comparing Stern-Gerlach devices with Polarizing beamsplitters}

The question then arising is, where this rejection occurs and why.
The purpose of this article is not to discuss in details the
likeliness of some possible explanations, but to focus on the
observable consequences of either fair or unfair sampling. We
would like nevertheless to make few remarks at this point.

By order of magnitude, the most important places for this
rejection are the detectors, since for the experiments performed
with photons, their efficiency was at best 10\%{}. The important
aspect is however not the magnitude of rejections, but the
selectiveness of these rejections. For this purpose, the analyzers
providing two-channel measurement are much better candidates
\cite{Hofer}. Indeed the experiments carried out so far were not
performed with Stern-Gerlach devices and atoms, but with
polarizing beamsplitters and photons. While Stern-Gerlach devices
can in principle perform an ideal two-channel measurement on all
emitted atoms, as the separation of the beam of atoms in two
channels occurs in a vacuum, the separation of the photons into
two channels by a polarizing beamsplitter occurs in a solid, i.e.
two birefringent prisms assembled together. Therefore, assuming
that a polarizing beamsplitter is an ideal analyzer similar to a
Stern-Gerlach device is a risky assumption. It would mean that the
polarizing beamsplitter treats all impinging photons equally,
providing a clear \(+\)1 or -1 as a result, while a 0 would occur
independently of the internal polarization of the photon. Even
though this problem was acknowledge on some rare occasions
\cite{Garra}, it was obviously not considered serious enough to
prevent the possibility of an experimental test using these
polarizing beamsplitters. Nevertheless, in our view this
assumption of fair sampling cannot {\itshape a priori} be accepted
as reasonable for two-channel polarizing beamsplitters. In other
words, our hypothesis is that on the contrary a polarizing
beamsplitter is {\itshape unfair} and that this behavior is
actually responsible for the observed EPR-Bell correlations
\footnote{We can think of few possible explanations for such such
a behavior. For instance, the shaky photon might have a greater
probability than other photons to be absorbed inside the
polarizing cube. Another possibility is that the state of the
impinging shaky photon is modified in such a way that it has a
greater probability of remaining undetected than others. For
instance, if a photon behaves in the polarizing beamsplitter more
like an electromagnetic wave than like a particle \cite{Lamb}, it
would split into two waves with reduces intensities, and if the
probability to generate a detectable signal depends on the
amplitude of the wave, then shaky photons would yield the least
detection probability.}.

\section{\label{sec:level5}Testing the fair sampling assumption for local hidden variable models}

Fair sampling is not only reputed to be a reasonable assumption,
which as we have seen above can be criticized \footnote{Not to
mention that attempting to justify fair sampling on the basis of
the symmetry of the experimental scheme \cite{Aspect82} is not a
convincing argument either, since our local realistic model
conforms exactly to the same requirements although it exhibits
unfair sampling.}, but it is also reputed impossible to test
experimentally. Yet, it is rather difficult to find in the
literature any clear justification for this line of thought. At
most some statements can be found relative to the one-channel type
experiment \cite{CHSH}, but to the best of our knowledge,
\emph{there is however no such a justification for the two-channel
type experiment}. We will actually show hereafter how fair
sampling can be put to a test, first with a known passive test for
which experimental investigations might unfortunately not be
conclusive, and then with a new active test that should provide
more contrasty results.

Let us assume that two sampling processes are at stakes in the
EPR-Bell two channel experiment, one fair, the other unfair. Note
that compared to the unfair sampling defined in Eq.
(\ref{unfairsampling}), a sampling process is fair if the
probability $P^r_{fair}$ for a particle to be rejected according
to this process does not depend on its hidden parameters (i.e.,
the internal polarization $\lambda$ in our model) and on the
measurement settings (i.e., the orientation $\varphi$ of the PBS).

Let us assume that we have at disposal a stable source of
entangled photons, so that the rate of particles emitted by the
source is a constant of time. Let $R$ be the total rate of pairs
entering in the coincidence circuitry. We assume that $R$ is
independent of the polarization distribution of the source and of
the measurement parameters $\varphi_A$ and $\varphi_B$, since the
source of entangled photons is rotationally invariant. Let
$R_{unfair}$ be the rate of pairs rejected according to the unfair
sampling process. This rate depends on the polarization
distribution $\rho$ of the source and on the orientations
$\varphi_A$ and $\varphi_B$ of the two-channel devices, that is,
$R_{unfair}=R_{unfair}(\rho,\varphi_A,\varphi_B)$ \footnote{The
fact that $R_{unfair}$ depends on $\rho$ doesn't mean in any way
that the sampling process itself is modified by the source that is
sent onto it. The sampling process characterizing the detection
pattern in Eq. (\ref{unfairsampling}) is independent of the
polarization distribution of the source $\rho$ that is sent onto
it: each particle is following the defined detection pattern of
Eq. (\ref{unfairsampling})---in our case
deterministically---independently of the polarization distribution
of the source it belongs to (i.e., there is no memory loophole).
In other words, the fairness or unfairness of the sampling
concerns the measurement setup, not the statistical properties of
the source that is sent onto it, and that precisely what makes it
possible to test experimentally.}. Finally, let $R_{fair}$ be the
rate of pairs rejected according to the fair sampling process,
which is on the contrary completely independent of these same
variables.

The experimentally available total rate of detected pairs in case
of unfair sampling can then be written as:
\begin{equation}\label{totalratedunfair}
    R_d(\rho,\varphi_A,\varphi_B)=R-R_{unfair}(\rho,\varphi_A,\varphi_B)-R_{fair}.
\end{equation}
The fair sampling assumption is then written as
$R_{unfair}(\rho,\varphi_A,\varphi_B)=0$, so that the total rate
of detected pairs $R_d$ is no more dependent on $\rho$,
$\varphi_A$, and $\varphi_B$ than are $R$ and $R_{fair}$:
\begin{equation}\label{totalratedfair}
    R_d(\rho,\varphi_A,\varphi_B)=R-R_{fair},\quad
    \forall\ \rho,\varphi_A,\varphi_B.
\end{equation}
Testing the fair sampling assumption therefore means to test which
one among equations (\ref{totalratedunfair}) and
(\ref{totalratedfair}) holds experimentally. Unlike with the
one-channel experiment, this can be decided experimentally by
making the settings $\rho$, $\varphi_A$, and $\varphi_B$ vary
\footnote{The trouble with the one-channel experiment that makes
checking of the fair sampling assumption impossible is that the
only measurement result that can be recorded is the +1 result
(channels -1 and 0 cannot be distinguished as they correspond both
to a non detection) so that the only available rate is $R_{++}$,
which always depends on $\rho$, $\varphi_A$, and $\varphi_B$,
whether or not the fair sampling assumption holds.}. In case of
unfair sampling, the measured total rate of detected pairs should
depend on these settings, according to Eq.
(\ref{totalratedunfair}), whereas it should remain independent of
them in case of fair sampling, according to Eq.
(\ref{totalratedfair}).

\subsection{\label{sec:level5a}A passive fair sampling test}

A first possible way to test the fair sampling assumption is to
vary the angle between the polarizing beamsplitters
$\Delta\varphi=|\varphi_B-\varphi_A|$, and check whether a
variation of \(R_{d}\) is observed.

\begin{figure}
\center
\includegraphics[width=7.0cm]{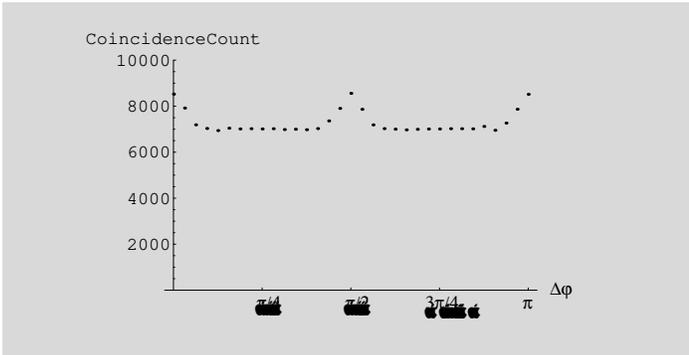}
\caption{\label{fig:epsart2} Total coincidence rate for the
unrealistic violation of CHSH-Bell inequality observed in
Fig.~\ref{fig:epsart1}. Unfair sampling and dependent errors
induce a sharp variation of \(R_{d}\).}
\end{figure}

We made a numerical simulation using the model described above
with settings exhibiting sharp correlations as in
Fig.~\ref{fig:epsart1}, and observed that the size of \(R_{d}\) is
not constant (see Fig.~\ref{fig:epsart2}) when $\Delta\varphi$ is
varied from 0 to \(\pi \).
This is due to the fact that the errors are dependent: if both
analyzers are oriented in the same direction modulo \(\pi \)/2,
then the chances that only one particle remains undetected are
smaller than for other relative angles, and since the total number
of undetected particles is a constant, the total number of
detected pairs is larger.
\begin{figure}
\center
\includegraphics[width=7.0cm]{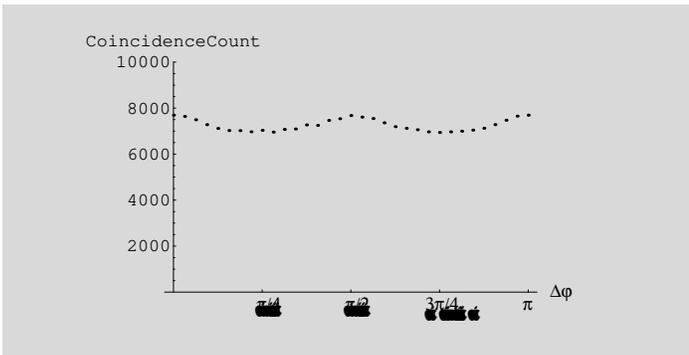}
\caption{\label{fig:epsart4} Total coincidence rate for the
realistic violation of CHSH-Bell inequality observed in
Fig.~\ref{fig:epsart3}. The variation of \(R_{d}\) appears less
sharply and might be easily mistaken for a constant in a real
experiment.}
\end{figure}
This effect might however be difficult to observe experimentally.
We have carried out a numerical simulation with the same loose
correlations that gave us a close fit with the predictions of
Quantum Mechanics, as in Fig.~\ref{fig:epsart2}, and found that
the error dependence is more difficult to observe (see
Fig.~\ref{fig:epsart4}), not to mention that in our model no dark
rates or no mistakes of any kind are implemented, which would
undoubtedly make this dependence even more difficult to observe,
so that checking fair sampling with this test can hardly be
conclusive \footnote{Note that an ad-hoc local realistic model
with independent error can be built \cite{Larsson2}, by defining
two distinct detector patterns for each particle in such a way
that the rate of detected particles remains independent of
$\Delta\varphi$ for the specific polarization distribution
$\rho^{\ast}$ of the source reproducing the EPR-Bohm statistics,
so that this passive test can only check that fair sampling is
reasonable, but cannot logically rule out unfair sampling.}.
\subsection{\label{sec:level5b}An active fair sampling test}
Nevertheless, another approach to testing fair sampling is
possible by varying not only $\varphi_A$ and $\varphi_B$ like in
the previous section, but also the polarization distribution
$\rho$ of the source, on which also depends $R_{unfair}$. Instead
of the rotationally invariant polarization distribution
$\rho^{\ast}$ of the source, we propose to use a source state with
probability distribution $\rho_\theta$ centered on a particular
polarization angle \(\theta \). We will see that varying
$\rho_\theta$, allows to exhibit clear cut discrepancies depending
on whether or not the fair sampling assumption holds \footnote{It
is perhaps necessary to stress that we are not in any way
suggesting that the source of entangled photons itself is not
rotationally invariant and that we want to put this to a test. We
do assume that the experimenter has checked before doing our test
that the source of entangled state is indeed rotationally
invariant, as it is clearly a crucial feature of the singlet
state, necessary to demonstrate any violation of a Bell
inequality. Our test is about breaking this rotationally invariant
on purpose to see how the measurement setup react to a source with
a preferred polarization.}.

\begin{figure}
\center
\includegraphics[width=8.0cm]{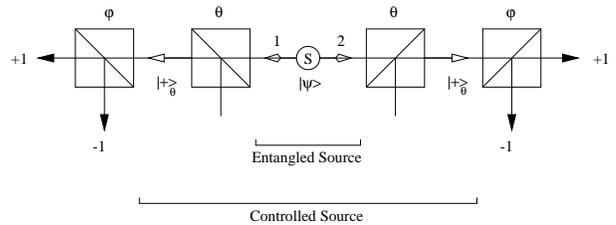}
\caption{\label{fig:epsart7} Scheme of controlled EPR-Bell
experiment to actively test fair sampling. The only difference
with the usual EPR-Bell experiment (see Fig.~\ref{fig:epsart8}) is
the source, which is controlled by the parameter $\theta$. In case
of unfair sampling, this setup should exhibit oscillations for the
total coincidence rate as in Fig.~\ref{fig:epsart6}.}
\end{figure}

In order to control the source experimentally, our proposal is
therefore to insert two aligned polarizing beamsplitters, both
oriented along the same angle \(\theta \), in the coincidence
circuitry right after the rotationally invariant source
$\rho^{\ast}$ of entangled photons and before the two-channel
measurement devices \footnote{We would like to stress that the
measurement setup itself is left unchanged with respect to the
ordinary EPR-Bell setup of Fig.~\ref{fig:epsart8}. In our fair
sampling test, represented in Fig.~\ref{fig:epsart7}, the one and
only modification concerns the source, and nothing else.}.

The procedure to simulate our active fair sampling test is the
following:
\begin{enumerate}
    \item
In all model based on the detection loophole, a detection pattern
for the couple PBS+Detector is given. It processes an
\emph{input}, dispatching each particle in the proper channel in
order to obtain the EPR-Bell statistics when a specific
rotationally invariant source $\rho^\ast$ impinges onto it. The
model usually says nothing as to the \emph{output} polarization
distribution of each channel, and that is what we need since our
test uses the output of a polarizer to control the polarization
distribution $\rho_\theta$ of the pairs.
    \item
This output polarization distribution must be set consistently
when impinging on the PBS+Detector pattern with the known behavior
of two successive polarizer, i.e., consistently with the Malus
law, and also consistently with the projection postulate of
Quantum Mechanics (that is, a polarizer does not only filter a
polarized beam, it also rotates its main polarization direction,
as can be seen by inserting a polarizer between two crossed
polarizers).
    \item
As described above, our active fair sampling test consists in
inserting two aligned polarizing beamsplitters oriented along an
angle \(\theta \) in the coincidence circuitry before the
two-channel measurement devices, both oriented along the same
angle \(\varphi \).
\end{enumerate}
\begin{figure}
\center
\includegraphics[width=7.0cm]{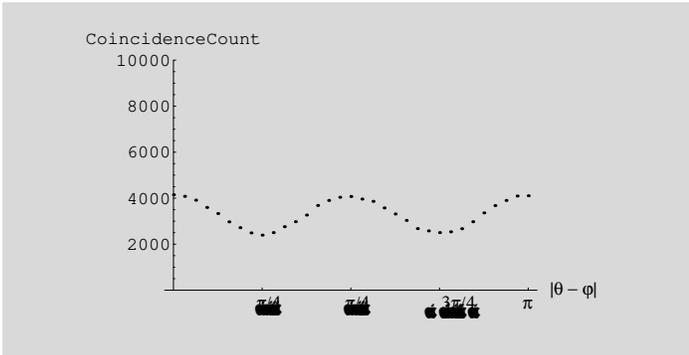}
\caption{\label{fig:epsart6} Total coincidence rate for a source
controlled with aligned polarizing beamsplitters at angle 0 and
for realistic settings (as in Fig.~\ref{fig:epsart3} and
Fig.~\ref{fig:epsart4}). The variation of \(R_{d}\) due to unfair
sampling appears sharply.}
\end{figure}
Note that this procedure is quite general and need not apply
solely to our model. Although local realist models based on the
detection loophole are always concerned only with reproducing the
EPR correlations (Step. 1), it is straightforward in most cases to
complete the model so as to get the Malus law (Step. 2), and
thereof perform our fair sampling test (Step. 3). We have actually
followed this procedure both with our model and the model that
seemed to be the least alike ours, that is Larsson's model, and
found that the output necessary to reproduce Step. 2 is the same,
and that the result of our test in Step. 3 shows exactly the same
behavior, as exhibited in Fig.~\ref{fig:epsart7}.

To be more specific, we gave a gaussian distribution to the
{\itshape output} polarization of the photons in the considered
channel (+1 in this case). In other words, the output state of
each particle dispatched in channel +1 was modified randomly
according to the context of the encountered polarizing
beamsplitter so that this output has statistically the form of a
gaussian distribution centered on the main polarization direction
of the polarizer \footnote{The gaussian distribution yielding a
good agreement with the Malus law has here a standard deviation of
$\pi/9$. The choice of a gaussian distribution is quite arbitrary,
as another distribution might give us a slightly better fit to the
Malus law, but whatever the output state it would just the same
have to be centered on a specific polarization $\theta$, with the
same qualitative result under our test.}. In a quantum
formulation, this modification of the state is nothing but the
collapsing of the initial vector state, from the rotationally
invariant singlet state to $|++\rangle_\theta$. By doing so we
have obtained the Malus law behavior without in any way modifying
our detection pattern for the EPR-Bell experiment. The numerical
simulation with the same settings as for the realistic violation
of Bell inequality of Fig.~\ref{fig:epsart2}, and \(\theta=0\),
shows a much clearer characteristic of unfair sampling (see
Fig.~\ref{fig:epsart6}), with very clear oscillations: the
contrast is roughly five times better than for the passive fair
sampling test of Fig.~\ref{fig:epsart4} (1/3 against 1/15). This
behavior can be explained in the following, all pairs coming from
the controlled source are such that $\lambda \simeq \theta$ modulo
$\pi$, so that if both analyzers are oriented in the same
direction \(\varphi_A=\varphi_B=\varphi \), the sum of all four
coincidence rates \({R_d}\) drops when $\varphi=\theta +\pi /4$
modulo $\pi /2$, because then almost all particles are shaky.

Thanks to its higher contrast, this behavior should be possible to
observe even with some noise and not so bright and accurate source
of entangled photons. It must be noted that these oscillations
cannot be made arbitrarily small without hindering the observed
EPR-Bell correlations: the higher the violation of Bell
Inequalities, the higher the oscillations of the total coincidence
rate in our fair sampling test.

Our test can therefore rule out fair sampling if these
oscillations are observed experimentally \footnote{Note that it is
possible to introduce an additional hidden-variable that tells
each photon whether the PBS should treat it according to a fair
sampling process or to an unfair one, depending on whether the
photon comes from the output of another polarizer of from the
output of a source of entangled photons. This would allow for the
model to reproduce both Malus law and EPR-Bell correlation, yet
with a constant $R_d$ under our test---we are grateful to
Jan-\r{A}ke Larsson for pointing out this idea. Although this
sounds definitely too \emph{ad hoc} for our taste, we must
acknowledge that it means that our test cannot logically rule out
unfair sampling. Since our test has never been performed, it is
nevertheless premature for now to try to thus hide a behavior that
might well be observed experimentally. We believe in any case that
nature itself cannot be that conspiratory in hiding unfair
sampling, so that our test can nevertheless tell whether or not
fair sampling is a reasonable assumption.}, since no fair sampling
process could account for such contrasty oscillations thus
connected to an apparent violation of Bell inequalities.

\section{\label{sec:level7}Conclusion}
The detection loophole still remains the most stringent of all
loopholes, so that the available EPR-Bell experiments would be far
more convincing if fair sampling was thoroughly investigated,
instead of assumed as being reasonable. The new experiment we have
proposed here should be simple to implement, and should allow to
check whether the fairness of the sampling in an EPR-Bell
experiments is a reasonable assumption.

\begin{acknowledgments}
We are most grateful to Jan-\r{A}ke Larsson for fruitful and
rather critical discussions on the issues raised in this article.
We are also indebted to Gregg Jaeger, Martin Salomon, Emilio
Santos, Afshin Shafiee and Johan Summhammer for their valuable
comments. This work was supported by the EU-Network ``Quantum
Probability and Applications".
\end{acknowledgments}
\bibliography{fairpra}
\end{document}